\documentclass{article}

\usepackage[final]{tackling_climate_workshop_style}

\usepackage[utf8]{inputenc} 
\usepackage[T1]{fontenc}    
\usepackage{hyperref}       
\usepackage{url}            
\usepackage{booktabs}       
\usepackage{amsfonts}       
\usepackage{amsmath}
\usepackage{amssymb}
\usepackage{nicefrac}       
\usepackage{microtype}      
\usepackage{booktabs}
\usepackage[dvipsnames]{xcolor}

\usepackage{graphicx}
\usepackage{authblk}
\usepackage{subcaption}

\title{Hybridizing Physics and Neural ODEs for Predicting Plasma Inductance Dynamics in Tokamak Fusion Reactors}

\author[1]{Allen M. Wang}
\author[1]{Darren T. Garnier}
\author[1]{Cristina Rea}
\affil[1]{MIT Plasma Science and Fusion Center, Cambridge, MA, USA}

\begin{document}

\maketitle

\begin{abstract}
While fusion reactors known as tokamaks hold promise as a firm energy source, advances in plasma control, and handling of events where control of plasmas is lost, are needed for them to be economical. A significant bottleneck towards applying more advanced control algorithms is the need for better plasma simulation, where both physics-based and data-driven approaches currently fall short. The former is bottle-necked by both computational cost and the difficulty of modelling plasmas, and the latter is bottle-necked by the relative paucity of data. To address this issue, this work applies the neural ordinary differential equations (ODE) framework to the problem of predicting a subset of plasma dynamics, namely the coupled plasma current and internal inductance dynamics. As the neural ODE framework allows for the natural inclusion of physics-based inductive biases, we train both physics-based and neural network models on data from the Alcator C-Mod fusion reactor and find that a model that combines physics-based equations with a neural ODE performs better than both existing physics-motivated ODEs and a pure neural ODE model.
\end{abstract}

\section{Introduction}
Recent advances in nuclear fusion have offered hope that it may arrive in time to help combat climate change by serving as a firm energy source to complement intermittent renewables. Within the fusion landscape, the tokamak is often considered the leading candidate for a fusion pilot plant, promising long-duration pulses with an existing physics basis offering a clear path to high energy gain \cite{wurzel2022progress, sorbom2015arc}. However, tokamaks are prone to sudden, highly energetic losses of plasma confinement induced by plasma instabilities. These events are known as ``plasma disruptions''. While plasma disruptions don't pose a threat to public safety, they do pose a significant threat to the economic viability of tokamaks \cite{maris2023impact}. This threat motivates the development of reliable high-performance real-time plasma control systems that can predict the onset of instabilities and ``soft-land'' the plasma by rapidly de-energizing it to a safe state.

One of the major challenges to realizing more reliable and high-performance tokamak plasma control is the difficulty of modelling plasma dynamics. While recent work successfully demonstrated the application of deep reinforcement learning to magnetic control of tokamak plasmas \cite{degrave2022magnetic}, the physics relevant to magnetic control, i.e. Ideal Magnetohydrodynamics (MHD), is the most well-simulated, and, furthermore, classical control techniques are already effective at tackling it.

To successfully control other aspects of plasma dynamics, advances in modelling are needed. In this regard, both purely physics-based and data-driven approaches currently fall short. On the purely physics-based side, the highest fidelity plasma simulations available today require millions of CPU hours to arrive at a steady state solution, making them unusable for control development, and even then require assumptions on boundary conditions such as the pressure gradient at the edge of the plasma \cite{rodriguez2022nonlinear}. This gap has motivated work on purely data-driven dynamics modelling with residual and recurrent neural networks, however, the relative paucity of data, distributional drift due to changes in hardware, and irregular measurement time-bases are major challenges \cite{char2023offline,abbate2021data}.

To tackle this problem, the neural ordinary differential equations (Neural ODEs) framework offers a promising path to improving the sample efficiency and robustness of data-driven prediction of plasma dynamics via the introduction of physics-based inductive biases. In this work, we explore the application of the neural ODE framework to a subset of the plasma dynamics problem. Namely, we model the coupled dynamics of plasma current, $I_p$, and plasma internal inductance $L_i$ to experimental data from the Alcator C-Mod fusion reactor. These coupled dynamics are of particular interest for ``soft landing'' as decreasing $I_p$ is one of the primary objectives of soft-landing, but doing so generally increases $L_i$, which is correlated with reduced plasma stability. Traditionally, partial differential equations (PDEs) would need to be evolved forward in time to predict the coupled dynamics of these two variables, but doing so is computationally intractable for the purposes of real-time control where decisions need to be made in milliseconds. Prior work \cite{romero2010plasma} derived a simple ODE model to predict these coupled dynamics. While parts of the derived model are high-confidence physics, parts of it involve intuition-based physics assumptions. In this work, we demonstrate that replacing the physics assumptions with a neural ODE, but keeping the high-confidence physics, yields a model that outperforms both the original model in \cite{romero2010plasma} and a full neural ODE.

\section{Background: The Romero Model}
\cite{romero2010plasma} derived a three state ODE system to model the coupled dynamics of $L_i$ and $I_p$. Of particular note is that two out of three equations are exact, with the third equation chosen based on physics intuition:
\begin{subequations}\label{eq:romero}
\begin{align}
    I_p\frac{dL_i}{dt} &= -2V_{ind} - 2V\\
    L_i\frac{dI_p}{dt} &= 2V_{ind} + V\\
    \frac{dV}{dt} &\approx -\frac{V}{\tau} - \frac{k}{\tau}V_{ind}\label{eq:romero_v_approx}
\end{align}
\end{subequations}
where $V_{ind}$ is the inductive loop voltage, a control variable that is easily controlled in real-time, $k, \tau$ are free parameters to be fit to data, and $V$ is a summary statistic of spatial variables that require evolving PDEs in time to calculate.

\section{Data}
We use a dataset of 489 shots, plasma-producing pulses of a fusion experiment, from the Alcator C-Mod fusion experiment. As internal inductance is not a directly measurable quantity, we used the estimated values from EFIT, a standard software tool in tokamaks used to infer the magnetic properties of a tokamak from sensor measurements \cite{lao1990equilibrium}. Of particular interest is that the measurement timebase is non-uniform; the time steps in our dataset have a mean of 19.416ms with a standard deviation of 3ms. While architectures such as recurrent neural networks (RNNs) do not naturally support non-uniform time bases, the Neural ODE framework easily handles this non-uniform timebase. We also performed a normalization of the state and control variables and introduced a constant factor where appropriate to the equations in \ref{eq:romero} to account for this normalization.

\section{Methodology}
\subsection{Neural ODEs with Control Inputs}
The classic neural ODE is given by:
\begin{align}
    \dot{\mathbf{x}} = f_\theta(\mathbf{x})
\end{align}
In our problem, we have the single control input $V_{ind}$ and thus we augment the standard Neural ODE with a control vector $\mathbf{u}$ to arrive at a system with the form:
\begin{align}
    \dot{\mathbf{x}} = f_\theta(\mathbf{x}, \mathbf{u})
\end{align}
All of our work was done in the Jax ecosystem, using Equinox to define models and Diffrax for differentiable ODE solvers \cite{kidger2021equinox,kidger2021on}. The control signal was interpolated in between samples with cubic Hermite splines with backwards differences, a method known for preserving causality \cite{morrill2021neural}.
\subsection{Multiple Shooting}
It has been observed in the literature that when training Neural ODEs with long time horizons, it is important to utilize ``multiple shooting'' to ensure stable and effective training \cite{massaroli2021differentiable}. In this work, we apply multiple shooting with a group size of 10 and zero continuity weight, as we found it was not necessary.
\subsection{Loss Function}
During training, each ``shot'' has a set of measurements $[\hat{\mathbf{x}}_0,...,\hat{\mathbf{x}}_N]$ where $\hat{\mathbf{x}}_i$ is sampled at time $t_i$, and where the sampling time intervals are irregular. We adopt the approach of utilizing an integral loss function to appropriately take into account this irregular time spacing. First, we define an ``instantaneous loss'' as such:
\begin{align}
    l(t) = \text{Huber}_{\delta=0.1}(\text{RelativeError}(\hat{\mathbf{x}}, \mathbf{x}_t))
\end{align}
Where we apply a Huber loss function with an outlier parameter of $0.1$ to the relative error to reduce the effect of outliers \cite{huber1992robust}. As we are interested in only predicting the coupled $L_i, I_p$ dynamics, we excluded the $V$ variable from this loss function. The loss function that is minimized is the time-integrated value of this instantaneous loss, which we approximate via the trapezoidal rule:
\begin{align}
    \mathcal{L} = \int_0^T l(t) dt \approx \sum_{i=0}^{T-1} \frac{(t_{i+1}-t_i)}{2}[l(t_{i+1}) - l(t_i)]
\end{align}
where $T$ is the total number of time steps of the episode.

\subsection{Models and Training}
In addition to \ref{eq:romero}, we train two more models: 1) a model that replaces equation \ref{eq:romero_v_approx} with a neural ODE mapping state and control to $\frac{dV}{dt}$, which we will refer to as ``RomeroNNV'',  and 2) a MLP that replaces all of the equations, a model we will refer to as ``MlpODE'': Both neural networks are fully connected MLPs with a width of 2, depth of 32, and softplus activation. Given the relatively small size of the dataset, the entire training set was included in each batch. We used the AdamW optimizer implemented in Optax with an exponential decay learning rate schedule starting at $5\times 10^{-3}$, a decay rate of 0.995 per epoch, and a terminal value of $1\times 10^{-4}$. The training was terminated if the mean validation loss over the past 1000 epochs had increased.

\section{Results}
\begin{figure}[!htb]
    \centering
    \includegraphics[width=\linewidth]{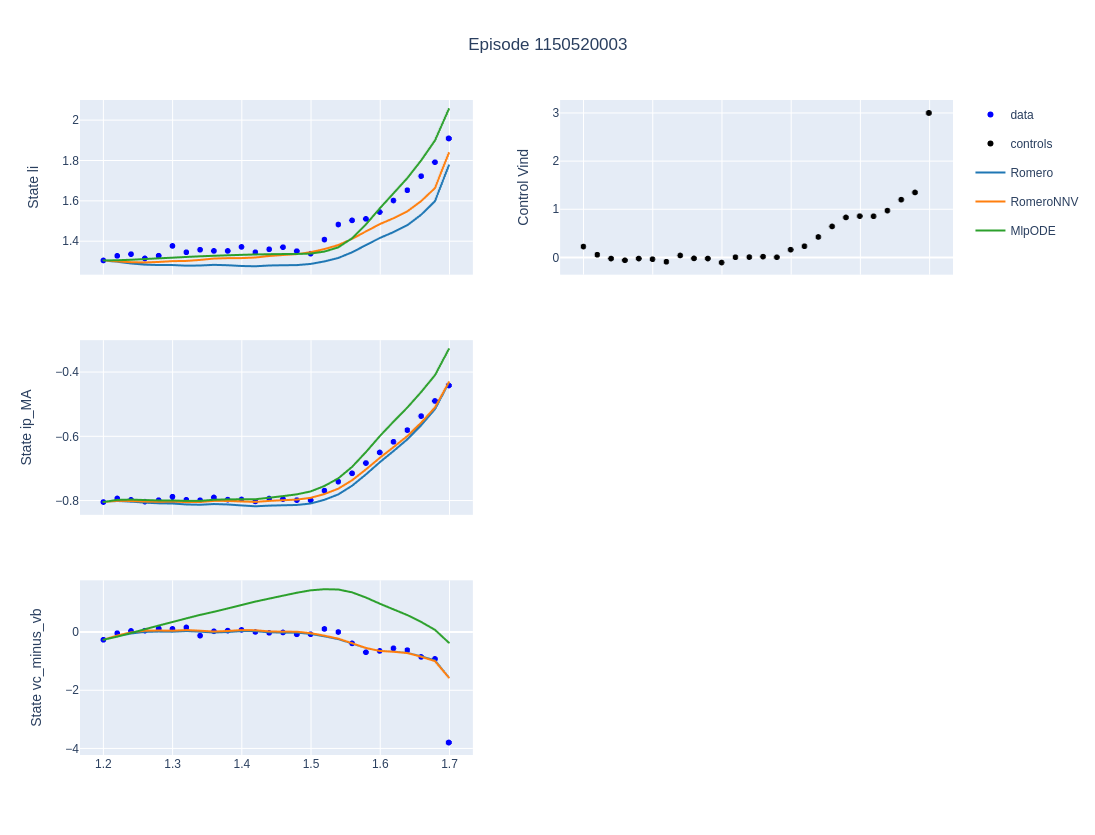}
    \caption{Example of model predictions against reactor data for Alcator C-Mod shot number 1150520003. Note that internal inductance is plotted with the standard tokamak normalization \cite{miyamoto2005plasma}, plasma current is in units of MA, and the $V$ variable is instead labeled as ``vc\_minus\_vb''}
    \label{fig:example_pred}
\end{figure}

While the training loss for the MlpODE continuously decreased and attained the lowest value out of the three models, the validation loss curves show it started over-fitting after several hundred epochs, a behavior not observed with the other two models, which have embedded physics structures. By both validation loss and test set accuracy metrics, shown in Table \ref{tab:test_metrics}, the RomeroNNV model performed the best with the MlpODE performing the worst on the test set. Figure \ref{fig:example_pred} shows an example of the three models' predictions against reactor data and control signals.
\begin{table}[ht]
    \centering
    \caption{Test set means and standard deviations of percent error at the end of each episode.}
    \label{tab:test_metrics}
    \begin{tabular}{c c c c}
        \toprule
        & \textbf{Romero} & \textbf{RomeroNNV} & \textbf{MlpODE} \\
        \midrule
        \( \text{li} \)   & $4.59\pm 4.37\%$ &  $\mathbf{3.50\pm3.70\%}$  &  $5.99\pm7.23\%$  \\
        \( \text{Ip} \)   &  $2.57\pm 2.34\%$  &  $\mathbf{1.97\pm1.90\%}$  &  $8.10\pm 9.69\%$ \\
        \bottomrule
    \end{tabular}
\end{table}
\begin{figure}[h]
    \centering
    \begin{subfigure}{0.425\textwidth}  
        \centering
        \includegraphics[width=\linewidth]{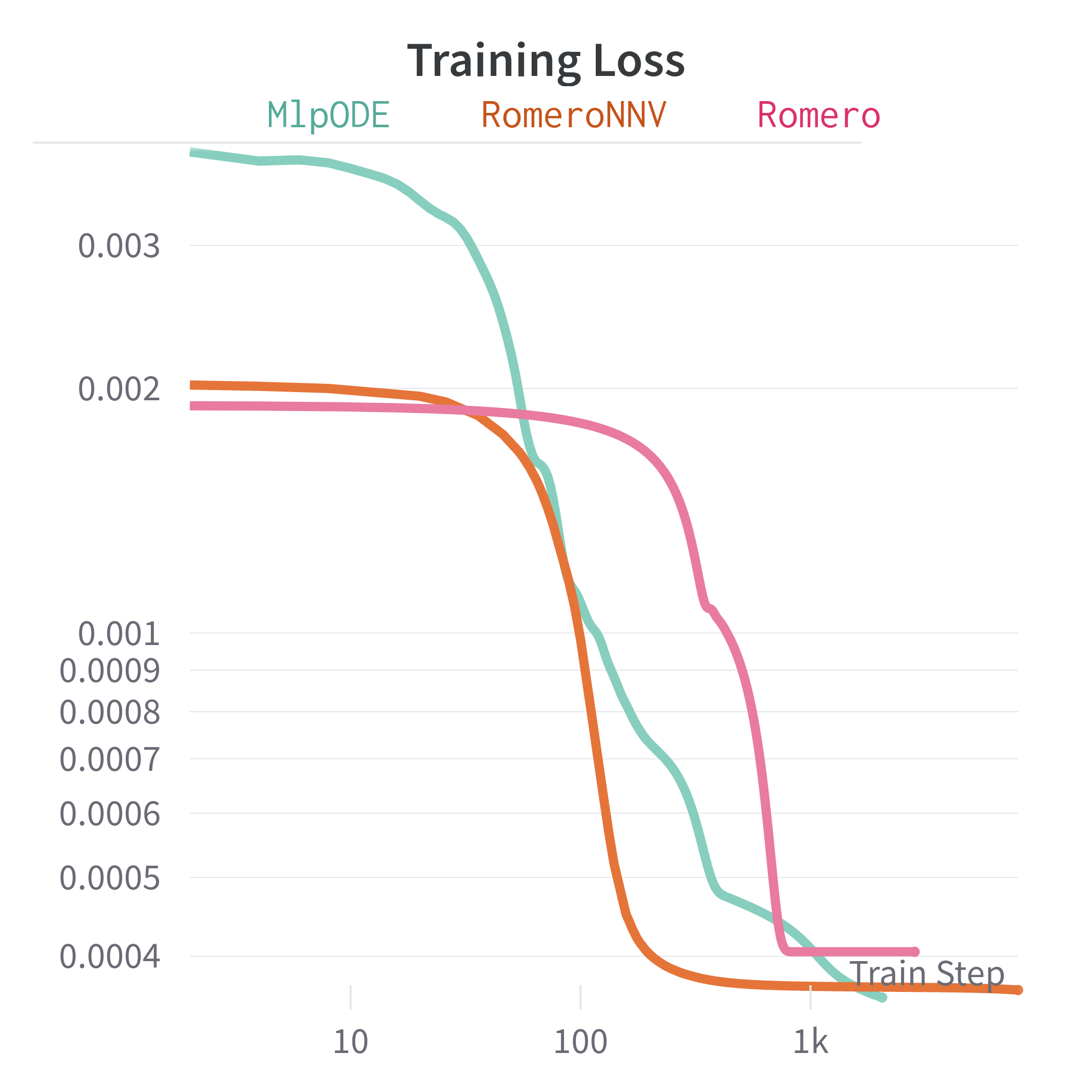}
    \end{subfigure}
    \begin{subfigure}{0.425\textwidth}  
        \centering
        \includegraphics[width=\linewidth]{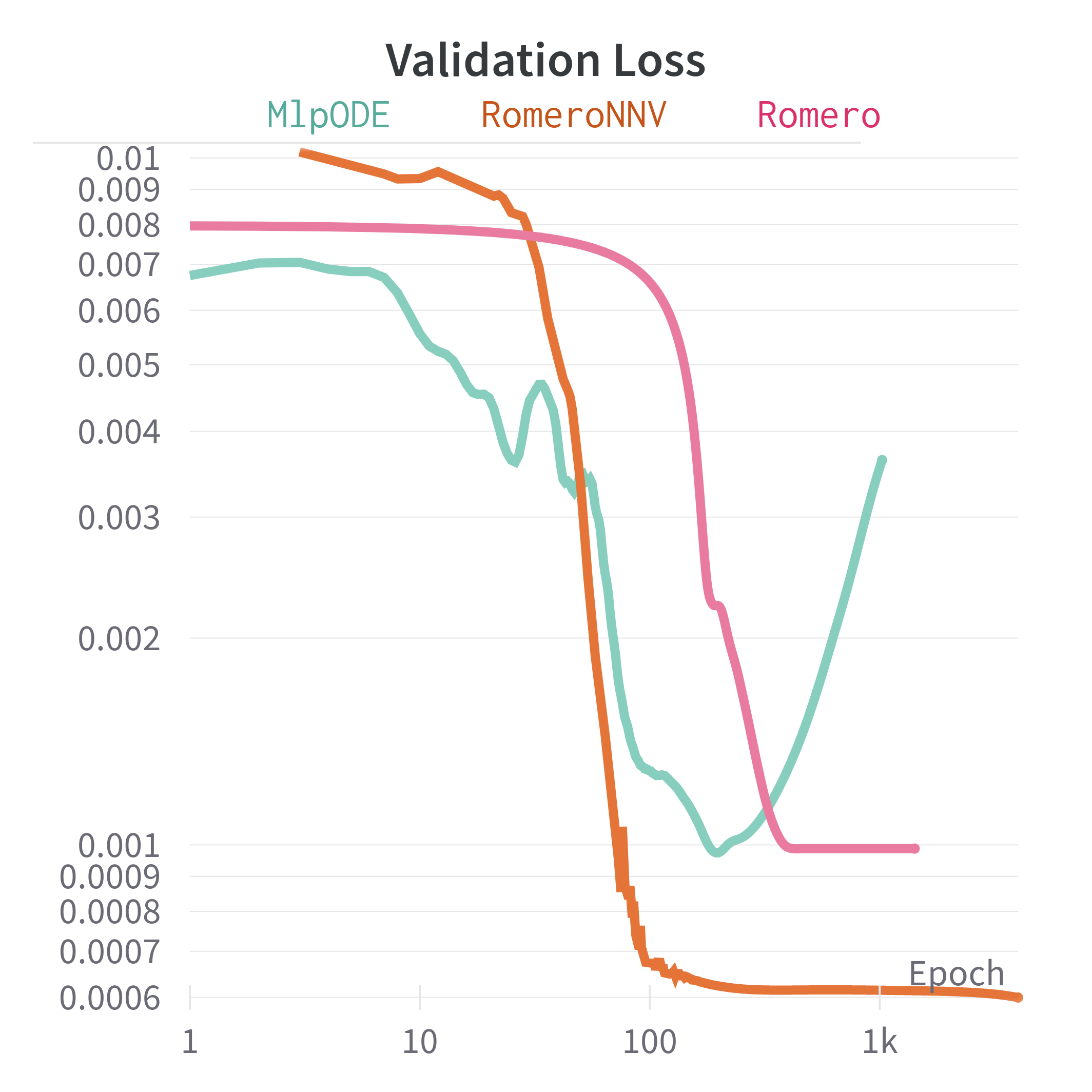}
    \end{subfigure}
    \caption{Training and validation losses for the three models.}
    \label{fig:train_val_dd}
\end{figure}
\subsection{Double Descent}
The increasing validation loss of the MlpODE model towards the end of the training run motivated longer runs of $3\times 10^4$ epochs to check for ``double descent'' behavior, which did appear \cite{nakkiran2021deep}. The run shown in Figure \ref{fig:train_val_dd} shows the MlpODE undergoing multiple rounds of overfitting followed by generalization. The RomeroNNV model proved to exhibit much more stable validation loss behavior, although it did appear to hit a plateau in both training and validation loss for thousands of epochs before descending further.
\begin{figure}[!h]
    \centering
    \begin{subfigure}{0.425\textwidth}  
        \centering
        \includegraphics[width=\linewidth]{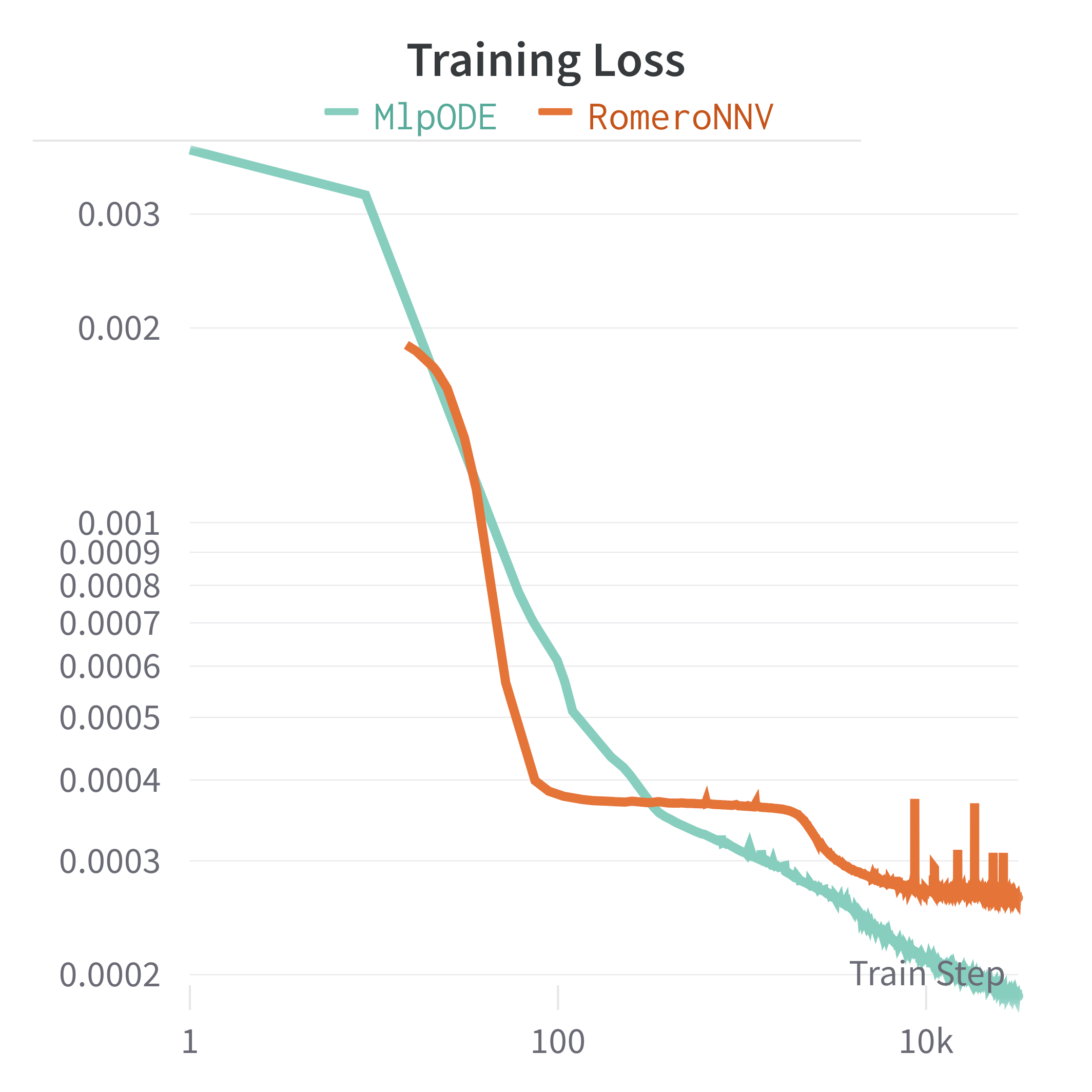}
    \end{subfigure}
    \begin{subfigure}{0.425\textwidth}  
        \centering
        \includegraphics[width=\linewidth]{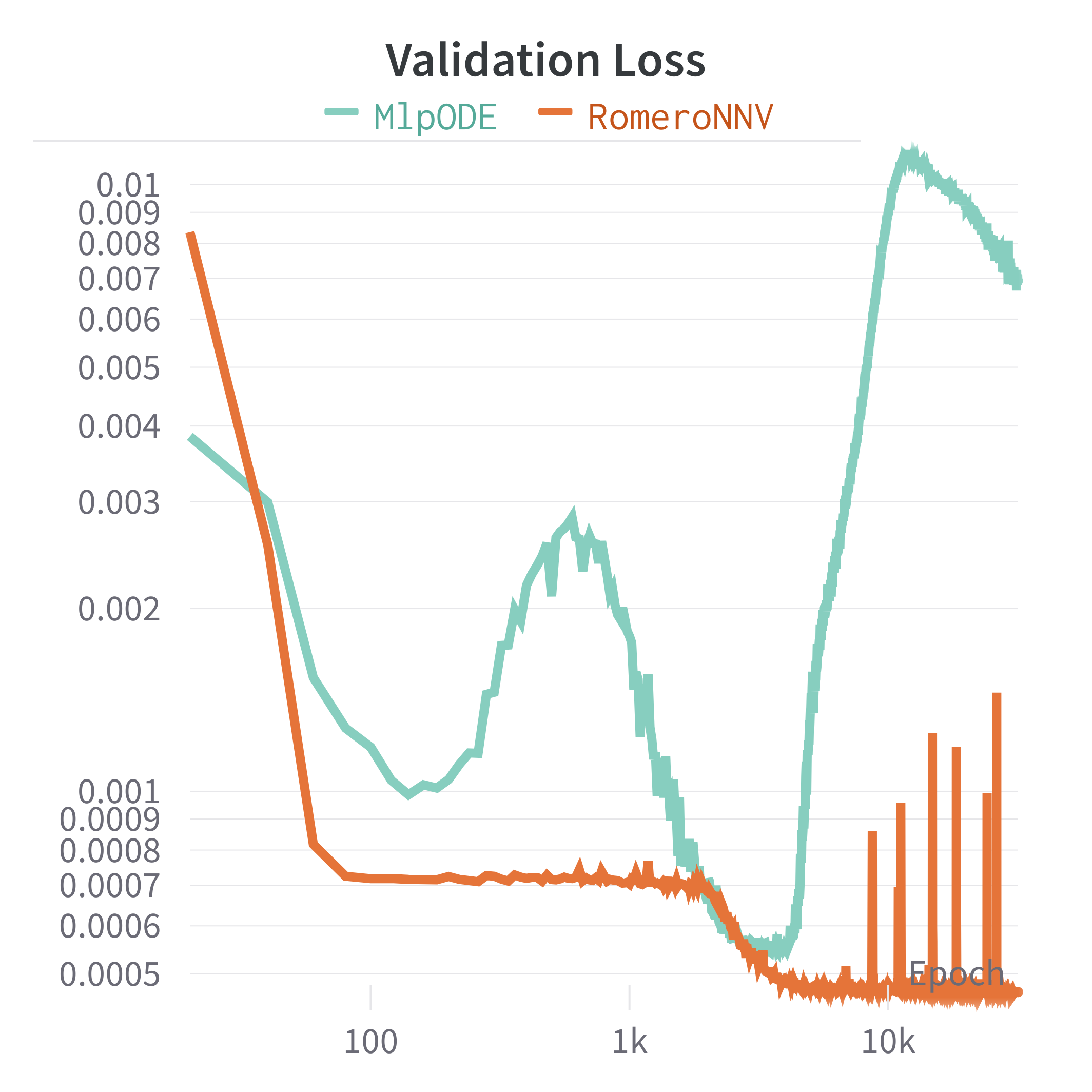}
    \end{subfigure}
    \caption{Training and validation loss from a longer training run of the two neural network models. Note that signs of double descent from shorter training runs motivated a more aggressive learning rate schedule starting at $7.5\times 10^{-3}$, ending at $10^{-3}$, with a decay rate of $0.9$ per every $500$ steps.}
    \label{fig:train_val_loss}
\end{figure}

\section{Conclusion}
We demonstrated the application of the neural ODE framework to predicting plasma dynamics and found that a hybrid dynamics model using both high-confidence physics paired with a neural network, which models the low confidence physics, outperforms the only known existing physics-motivated model, and a pure neural ODE. Future work should investigate adopting a similar modelling approach to a broader range of plasma dynamics.
\newpage
\section{Acknowledgements}

\bibliography{references}

\begin{thebibliography}{16}
\providecommand{\natexlab}[1]{#1}
\providecommand{\url}[1]{\texttt{#1}}
\expandafter\ifx\csname urlstyle\endcsname\relax
  \providecommand{\doi}[1]{doi: #1}\else
  \providecommand{\doi}{doi: \begingroup \urlstyle{rm}\Url}\fi

\bibitem[Abbate et~al.(2021)Abbate, Conlin, and Kolemen]{abbate2021data}
J.~Abbate, R.~Conlin, and E.~Kolemen.
\newblock Data-driven profile prediction for diii-d.
\newblock \emph{Nuclear Fusion}, 61\penalty0 (4):\penalty0 046027, 2021.

\bibitem[Char et~al.(2023)Char, Abbate, Bard{\'o}czi, Boyer, Chung, Conlin,
  Erickson, Mehta, Richner, Kolemen, et~al.]{char2023offline}
I.~Char, J.~Abbate, L.~Bard{\'o}czi, M.~Boyer, Y.~Chung, R.~Conlin,
  K.~Erickson, V.~Mehta, N.~Richner, E.~Kolemen, et~al.
\newblock Offline model-based reinforcement learning for tokamak control.
\newblock In \emph{Learning for Dynamics and Control Conference}, pages
  1357--1372. PMLR, 2023.

\bibitem[Degrave et~al.(2022)Degrave, Felici, Buchli, Neunert, Tracey,
  Carpanese, Ewalds, Hafner, Abdolmaleki, de~Las~Casas,
  et~al.]{degrave2022magnetic}
J.~Degrave, F.~Felici, J.~Buchli, M.~Neunert, B.~Tracey, F.~Carpanese,
  T.~Ewalds, R.~Hafner, A.~Abdolmaleki, D.~de~Las~Casas, et~al.
\newblock Magnetic control of tokamak plasmas through deep reinforcement
  learning.
\newblock \emph{Nature}, 602\penalty0 (7897):\penalty0 414--419, 2022.

\bibitem[Huber(1992)]{huber1992robust}
P.~J. Huber.
\newblock Robust estimation of a location parameter.
\newblock In \emph{Breakthroughs in statistics: Methodology and distribution},
  pages 492--518. Springer, 1992.

\bibitem[Kidger(2021)]{kidger2021on}
P.~Kidger.
\newblock \emph{{O}n {N}eural {D}ifferential {E}quations}.
\newblock PhD thesis, University of Oxford, 2021.

\bibitem[Kidger and Garcia(2021)]{kidger2021equinox}
P.~Kidger and C.~Garcia.
\newblock {E}quinox: neural networks in {JAX} via callable {P}y{T}rees and
  filtered transformations.
\newblock \emph{Differentiable Programming workshop at Neural Information
  Processing Systems 2021}, 2021.

\bibitem[Lao et~al.(1990)Lao, Ferron, Groebner, Howl, John, Strait, and
  Taylor]{lao1990equilibrium}
L.~Lao, J.~Ferron, R.~Groebner, W.~Howl, H.~S. John, E.~Strait, and T.~Taylor.
\newblock Equilibrium analysis of current profiles in tokamaks.
\newblock \emph{Nuclear Fusion}, 30\penalty0 (6):\penalty0 1035, 1990.

\bibitem[Maris et~al.(2023)Maris, Wang, Rea, Granetz, and
  Marmar]{maris2023impact}
A.~D. Maris, A.~Wang, C.~Rea, R.~Granetz, and E.~Marmar.
\newblock The impact of disruptions on the economics of a tokamak power plant.
\newblock \emph{Fusion Science and Technology}, pages 1--17, 2023.

\bibitem[Massaroli et~al.(2021)Massaroli, Poli, Sonoda, Suzuki, Park,
  Yamashita, and Asama]{massaroli2021differentiable}
S.~Massaroli, M.~Poli, S.~Sonoda, T.~Suzuki, J.~Park, A.~Yamashita, and
  H.~Asama.
\newblock Differentiable multiple shooting layers.
\newblock \emph{Advances in Neural Information Processing Systems},
  34:\penalty0 16532--16544, 2021.

\bibitem[Miyamoto(2005)]{miyamoto2005plasma}
K.~Miyamoto.
\newblock \emph{Plasma physics and controlled nuclear fusion}, volume~38.
\newblock Springer Science \& Business Media, 2005.

\bibitem[Morrill et~al.(2021)Morrill, Kidger, Yang, and
  Lyons]{morrill2021neural}
J.~Morrill, P.~Kidger, L.~Yang, and T.~Lyons.
\newblock Neural controlled differential equations for online prediction tasks.
\newblock \emph{arXiv preprint arXiv:2106.11028}, 2021.

\bibitem[Nakkiran et~al.(2021)Nakkiran, Kaplun, Bansal, Yang, Barak, and
  Sutskever]{nakkiran2021deep}
P.~Nakkiran, G.~Kaplun, Y.~Bansal, T.~Yang, B.~Barak, and I.~Sutskever.
\newblock Deep double descent: Where bigger models and more data hurt.
\newblock \emph{Journal of Statistical Mechanics: Theory and Experiment},
  2021\penalty0 (12):\penalty0 124003, 2021.

\bibitem[Rodriguez-Fernandez et~al.(2022)Rodriguez-Fernandez, Howard, and
  Candy]{rodriguez2022nonlinear}
P.~Rodriguez-Fernandez, N.~Howard, and J.~Candy.
\newblock Nonlinear gyrokinetic predictions of sparc burning plasma profiles
  enabled by surrogate modeling.
\newblock \emph{Nuclear Fusion}, 62\penalty0 (7):\penalty0 076036, 2022.

\bibitem[Romero et~al.(2010)Romero, Contributors, et~al.]{romero2010plasma}
J.~Romero, J.-E. Contributors, et~al.
\newblock Plasma internal inductance dynamics in a tokamak.
\newblock \emph{Nuclear Fusion}, 50\penalty0 (11):\penalty0 115002, 2010.

\bibitem[Sorbom et~al.(2015)Sorbom, Ball, Palmer, Mangiarotti, Sierchio,
  Bonoli, Kasten, Sutherland, Barnard, Haakonsen, et~al.]{sorbom2015arc}
B.~Sorbom, J.~Ball, T.~Palmer, F.~Mangiarotti, J.~Sierchio, P.~Bonoli,
  C.~Kasten, D.~Sutherland, H.~Barnard, C.~Haakonsen, et~al.
\newblock Arc: A compact, high-field, fusion nuclear science facility and
  demonstration power plant with demountable magnets.
\newblock \emph{Fusion Engineering and Design}, 100:\penalty0 378--405, 2015.

\bibitem[Wurzel and Hsu(2022)]{wurzel2022progress}
S.~E. Wurzel and S.~C. Hsu.
\newblock Progress toward fusion energy breakeven and gain as measured against
  the lawson criterion.
\newblock \emph{Physics of Plasmas}, 29\penalty0 (6), 2022.

\end{thebibliography}

\newpage
\section*{Supplementary Material}

\end{document}